\documentstyle[12pt,epsf,psfig]{article}
\input epsf
\def\beq{\begin{equation}}
\def\eeq#1{\label{#1}\end{equation}}
\def\eeqn{\end{equation}}
\def\beqa{\begin{eqnarray}}
\def\eeqa#1{\label{#1}\end{eqnarray}}
\def\eeqan{\end{eqnarray}}

\begin{document}
\begin{flushright}
NYU-TH/01/08/07 \\
\end{flushright}

\vspace{0.1in}
\begin{center}
\bigskip\bigskip
{\large \bf Changing $\alpha$ With Time: Implications For
Fifth-Force-Type Experiments And Quintessence}

\vspace{0.5in}      

{Gia Dvali and Matias Zaldarriaga}
\vspace{0.1in}

{\baselineskip=14pt \it 
Department of Physics, New York University,  
New York, NY 10003\\}
\vspace{0.2in}
\end{center}

\vspace{0.9cm}
\begin{center}
{\bf Abstract}
\end{center} 
\vspace{0.1cm}
If the recent observations suggesting a time variation of the fine
structure constant are correct, they imply the existence of an ultra
light scalar particle. This particle inevitably couples to nucleons
through the $\alpha$-dependence of their masses and thus mediates an
isotope-dependent long-range force. The strength of the coupling is
within a couple of orders of magnitude from the existing experimental
bounds for such forces.  The new force can be potentially measured in
the precision experimental tests of the equivalence principle.  Due to
an interesting coincidence of the required time-scales, the scalar
filed in question can at the same time play the role of a quintessence
field.

\vspace{0.1in}

\newpage

\section{Introduction}

A number of recent observations of absorption lines in high redshift 
QSOs suggest the time variation of the fine
structure constant, $\alpha$ over the cosmological time scales.  The
favored value of the change ${\Delta\alpha \over \alpha} \simeq -0.72
\times 10^{-5}$ over the redshift range $0.5 < z < 3.5$ \cite{alpha}.
If true, this striking effect can be interpreted as a signal of the
new physics beyond the standard model.  In such a situation it is
important to understand other possible observable consequences and
experimental tests of this phenomenon.
 
In the present note we argue that time variation of $\alpha$ implies
the existence of a very weakly-coupled ultra light scalar particle
$\phi$ (the ``$\alpha$-ion'').  The $\alpha$-ion
necessarily couples to ordinary
nucleons, protons and neutrons, through the $\alpha$-dependence of
their masses.  Thus $\phi$ mediates a {\it composition-dependent}
``fifth force''-type long range interaction. This type of interactions
are subject to experimental constraints such as tests coming from
searches for violations of the equivalence principle.  The time-scale
and strength of suggested $\alpha$-variation implies that the strength
of the coupling is close to the existing experimental limits and may
be tested in future experiments with improved precision.

In the following sections we discuss our argument 
in more detail.

\section{Fifth Force from Changing $\alpha$}

The standard picture of our Universe assumes that at macroscopic
length scales all the way up to the present Hubble size $\sim
10^{28}$cm nature is described by an effective {\it four-dimensional}
low energy field theory. Although this is not the only possibility
(for instance extra dimensions may open up at some astronomical
scales), we will adopt this standard picture throughout the present
discussion. 

In the effective four-dimensional field theory the only consistent way
known to make Lagrangian parameters time dependent is through
promoting them into functions of some dynamical order parameter, an
elementary (or composite) scalar field.  Perhaps the most well known
examples of this sort occur in string theory where the gauge and the
gravitational coupling ``constants'' are set by the expectation values
of the scalar fields such as the dilaton or the string moduli.  Thus,
the time variation of $\alpha$ within a 4D field theory {\it
necessarily} implies that $\alpha$ is a function of a time-dependent
scalar field $\phi$. The unusual thing about $\phi$ is that to cause a
change in $\alpha$ during the last Hubble time it should be
extraordinary light, with mass comparable to the present Hubble
scale $H\sim 10^{-33}$eV. This follows from the equation of motion for
a scalar field of mass $m$ in the expanding Universe,
\begin{equation}
\ddot \phi+ 3 H \dot \phi + m^2 \phi +... = 0.
\end{equation}
If the mass term is much smaller that the Hubble scale then the
friction term in the equation dominates, the field does not move and
so it is unable to produce the required change in $\alpha$.  If
however the mass is much larger that the present Hubble scale, the
field starts to oscillate much earlier in the history of the Universe
and would now be at the minimum of the potential. Thus the only way to
have a field changing at the present time is if its mass is of order
the current Hubble scale.

The most general expansion of
the function $\alpha(\phi)$ about its present day value $\alpha_0=
\alpha(\phi=\phi_{today})$ can be written as 
\beq \alpha = \alpha_0 +
\lambda {\phi \over M_P} +...
\label{coupling}
\eeq
where $M_P$ is the Planck mass and $\lambda$ is some constant. 
We shall assume the absence of any
fine tunning among the different terms of expansion. Under such an
assumption the observation that $\Delta\alpha/\alpha\sim 10^{-5}$ 
suggests that $\lambda {\Delta\phi \over M_P}\sim 10^{-7}$
within the last Hubble time. Assuming for the moment that
$\Delta\phi \sim M_P$ or smaller, we get $\lambda > 10^{-7}$.

Let us now discuss the experimental constraints on $\lambda$. These
constraints come from the fact that $\phi$ inevitably couples to
protons and neutrons and thus, being so light, 
should mediate a long-range force. The
coupling to nucleons follows from the electromagnetic corrections to
the nucleon mass.  To leading order in $\alpha$ these corrections
can be written as\cite{gasser}:
\begin{equation}
\delta m_p = B_p \alpha = 0.63 MeV~~~\delta m_n = B_n \alpha = -0.13 MeV
\end{equation}
where $m_n$ and $m_p$ are the neutron and the proton masses respectively.
Thus through the dependence on $\alpha$ the nucleon masses are promoted into
the functions of $\phi$. Nucleons-nucleon-$\phi$ couplings can be easily
read-off by expanding the nucleon mass in powers of $\phi$ 
in the effective low energy Lagrangian
\begin{equation}
L= m_N(\phi)\bar{N}N +...
\end{equation}
where $N$ stands for either neutron or a proton. The resulting leading order
couplings are
\begin{equation}
    {B_N\lambda \over M_P}\phi\bar{N}N, ~~~N=p,n
\end{equation}
Thus, exchange of $\phi$ leads to a long range force. Due to the
difference in the couplings to neutron and proton, the force in
question is isotope-dependent and can lead to an apparent violation of
the equivalence principle.

A non-relativistic test body
of inertial mass $m$ placed in the gravitational
field of earth at distance $r$ from the center will
undergo the following acceleration,
\begin{equation}
a = a_{gr} + a_{\phi}
\end{equation}
where $a_{rg}={M_E/M_p^2 r^2}$ is the usual Newtonian
acceleration and
\begin{equation}
a_{\phi} = {1\over r^2}{\lambda^2 \over M_P^2 m}(n_n^EB_n + 
n_p^EB_P)(n_nB_n + n_pB_P)
\end{equation}
is the acceleration induced by the $\phi$-force. Here and $n_{n,p}^E$ and
 $n_{n,p}$ are numbers of protons and neutrons in the
the Earth and in the test body respectively.
In this estimate we ignore the small correction
due to ``off-shellness'' of nucleons inside the nuclei and atoms
coming from the binding energy.
 
The difference in accelerations
between the two bodies can be measured in E\"otv\"os-type experiments
(for a review see, e.g., \cite{will}).
The convenient parameter is the so-called ``E\"otv\"os ratio''
\begin{equation}
\eta = {2|a_1 - a_2|\over |a_1 + a_2|}
\end{equation}
where $a_1$ and $a_2$ are the accelerations of two different bodies.
In the present case this parameter is given by
\begin{equation}
\eta = {\Delta a_{\phi}\over a_{gr}} = 
{\lambda^2 \over \bar{m}_{N}^2}(f_n^EB_n + 
f_p^EB_P)(\Delta f_nB_n + \Delta f_pB_P)
\end{equation}
where $\bar{m}_{N}$ is an
average nucleon mass which we take to be $ 1$GeV, and $f_{n,p}^E$ and
 $f_{n,p}$ are the average fractional numbers of protons and neutrons in the
nuclei of the Earth and of the test bodies respectively. 
$\Delta f_{n,p}$ is the difference of these quantities between two different
accelerated bodies. We will take
$f_{n,p}^E \approx 1/2$ 
for our estimates. For typical materials used in the experiments 
such as copper ($f_{p}\sim 0.456$), lead ($f_{p}\sim 0.397$)
or uranium ($f_{p}\sim 0.385$) ,
we have  $\Delta f_{n,p}\approx 6\times 10^{-2}\sim 10^{-1}$.
We thus estimate, 
\begin{equation}\label{eta}
\eta \approx  10^{-3} \lambda^2 
\end{equation}

Equation (\ref{eta}) and the present day experimental bound
$\eta < 10^{-13}$ coming from E\"ot-Was\cite{Eot-Was} measurements
give the following bound,
\begin{equation}
 \lambda < 10^{-5} 
\end{equation}
This is in no contradiction with a fractional variation of $\alpha$ at
the $10^{-5}$ level provided $\phi$ changed by more than 
$\Delta\phi > 10^{-2}M_P$ during the last
Hubble period. It is interesting that the suggested value of
$\lambda$ for maximal variations ($\Delta\phi\sim M_P$) 
is only within couple of orders of magnitude from the experimental
limit so it can be potentially observed in future measurements with
improved precision. To put this necessary increase in sensitivity in
context, the constrains on $\eta$ have
improved almost two orders of magnitude during the last decade.

One may wonder how stringent the lower bound on $\lambda$ obtained
from the observed variation of $\alpha$ 
assuming $\Delta\phi/M_P \sim 1$ really is. 
Naively $\lambda$ could be arbitrarily small, since even with small
$\lambda$ one can still make-up for an observed variation of $\alpha$
by assuming that $\Delta\phi >> M_P$ per Hubble period.
However, large changes in the field are difficult to accommodate.

%Namely $\Delta\phi \sim M_P$ implies the upper limit for variation
%of $any$ scalar field during the last Hubble time, implying that
%it is unreasonable to expect $\lambda < 10^{-7}$.
% To prove this 
Let us assume that $\Delta\phi >>M_P$ during
the last $\delta t \sim H^{-1}$ and show that this assumption
leads us to an inconsistency. Indeed, such a fast late-time variation of $\phi$
would imply that
\begin{equation}
 {\Delta\phi \over H^{-1}} >> M_PH
\end{equation}
Thus, the average kinetic energy of the field during this period is
\begin{equation}
\rho_{kin}= {1\over 2}\left (
{d\phi \over dt}\right )^2 \sim \left ({\Delta\phi \over H^{-1}}
\right )^2 >> M_P^2H^2
= \rho
\end{equation}
where $\rho$ is the total energy density. Thus, the kinetic energy of
$\phi$ has to be larger than the total energy density of the Universe,
which is impossible.

Equivalently we can write,
\begin{equation}\label{rhok}
\rho_{kin} = {1\over 2}\left (
{d\phi \over dt} \right )^2 = (1+\omega_\phi)\rho_\phi
\end{equation}
where $\omega_\phi$ is the equation of state parameter for the filed
$\phi$ ($\omega_\phi=p_\phi/\rho_\phi$). Equation (\ref{rhok}) leads to:
\begin{equation}
\left ({\Delta\phi \over M_P} \right )^2 \sim 
(1+\omega_\phi){\rho_\phi \over \rho}. 
\end{equation}
There are clearly two possibilities. First if the field $\phi$
dominates the energy density of the Universe today then $\Delta\phi$
can be of the order of $M_P$ (although slightly smaller if the
Universe is accelerating as implied by recent observations of high
redshift supernovae which imply $\omega_\phi < -0.6$ \cite{super}) and then
$\lambda \sim 10^{-7}$.  In this case there will be clear signatures
of the existence of $\phi$ in the data from future astronomical observations
(for a detailed forecast of the ability of future supernovae
experiments to measure the cosmic equation of state see \cite{astier};
see \cite{tegmark} for a summary of what constraints on the time
evolution of the energy density of the Universe can be expected
from different types of astronomical observations planned or
under way). The second possibility is that the energy density in
$\phi$ is sub-dominant today. In that case $\Delta\phi << M_P$ and so
$\lambda >> 10^{-7}$ and thus should be near the current capability of
E\"otv\"os-type experiments.

The implied properties of a field that can produce the observed 
variation of $\alpha$ are such that it should also manifest itself as
either as a quintessence\cite{quin} field detectable by future astronomical
observations or in future precision tests of the equivalence
principle. 

\section{Discussions}

We have argued that the time-variation of $\alpha$ implies the
existence of a super-light scalar particle, which would inevitably
couple to protons and neutrons through the $\alpha$-dependence of their
masses. Thus it mediates a  potentially measurable isotope-dependent
force. Let us briefly discuss some loopholes in our
arguments.

{\bf Symmetry protection.} One may argue that the expansion
(\ref{coupling}) may not include the linear term and start from a
higher power of $\phi$.  This may be achieved by postulating the
symmetry $\phi \rightarrow -\phi$.  However, since $\phi$ is a {\it
time-dependent} field changing throughout the history of the Universe,
such a symmetry must be inevitably broken during most of the history.
Absence of the linear term today would imply that we happen to live at
a very special point of the restored symmetry. However, this would
require a miraculous coincidence, since $\phi$ is an extremely slowly
changing field on the Hubble scales and has ``spent'' most of the time
away from the restored symmetry point.  There is no reason whatsoever 
for such a mode to approach the enhanced symmetry point precisely at
the time when the fifth force measuring experiments are taking place
on Earth.  Such a coincidence is very unlikely and we disregard it.

{\bf Breakdown of 4D effective field theory.} Our results relied on
four-dimensional effective field theory arguments, which requires
that the time dependence of the effective parameters should occur 
through their dependence on time-dependent fields.  It is conceivable
however that the Universe is not four-dimensional at large
distances in which case our arguments could break down. For instance,
there are models in which gravity and electromagnetic
interactions become five-dimensional at Hubble scales\cite{dgp}.
It is possible that in these
scenarios the variation of electromagnetic and gravitational constants
may not necessarily be related with the existence of a
four-dimensional scalar field mediating $1/r^2$ force, but rather with
some high-dimensional mode mediating a much weaker force.

 {\bf Universally coupled $\phi$.} One could imagine that $\phi$-dependence
is also experienced by other parameters of the standard model (such as,
for instance gravitational constant, $\alpha_{strong}$ and the fermion masses)
in such a way that the net non-universal
effects in proton and neutron couplings partially
cancel out. In such a case the bounds on $\alpha$-ion
couplings would be somewhat milder. Such a conspiracy is hard to achieve
and we have ignored this possibility in the present work.

 {\bf $\alpha$-variation in stars.} Finally, let us
note some other possible observational consequences of $\alpha$-ion.
Due to it's extremely small mass the value of $\phi$ can
significantly change
in a strong electromagnetic field or in a dense medium, e.g. such as the
vicinity of a neutron star. For instance, in a background neutron
density $\phi$ acquires an effective potential, which in the leading order
is $V(\phi)\sim \alpha(\phi)B_nN_{neutron}$, where $N_{neutron}$
is the number density. This potential forces $\phi$ to depart
from its average value locally and may lead to an observable change
in $\alpha$. 

$~~~~~~~~$

%\vspace{0.5cm} \\
{\bf Acknowledgments}:
We are grateful to Akaki Rusetsky for valuable comments and discussions
on low energy QCD. We would like to thank Andrei
Gruzinov, Arthur Lue, Massimo Porrati and Roman Scoccimarro for
useful discussion. The work of GD was supported in part by David
and Lucille Packard
Foundation Fellowship for Science and Engineering, by Alfred P. Sloan
foundation fellowship and by NSF grant PHY-0070787.


\begin{thebibliography}{99}

\bibitem{alpha}
J. K. Webb et al, [astro-ph/0012539].

\bibitem{gasser} J.~Gasser and H.~Leutwyler,
Phys. Rept.  {\bf 87}, 77 (1982).

\bibitem{will} C. M. will, [gr-qc/0103036].

\bibitem{Eot-Was}
S. Baessler et al, Phys. Rev. Lett. {\bf 83}, 3585 (1999).

\bibitem{super}
P. M. Garnavich et al, Astrophys. J. {\bf 509}, 74 (1998);
A. G. Riess et al, Astrophys. J. {\bf 116} 1009 (1998)
S. Perlmutter et al, [astro-ph/9812133]; 
A. G. Riess et al,[astro-ph/0104455].

\bibitem{astier}
M. Goliath, R. Amanullah, P. Astier, A. Goobar, R. Pain; 
[astro-ph/0104009].

\bibitem{tegmark}
M. Tegmark, [astro-ph/0101354].
\bibitem{quin}
For a review see, e.g., J. P. Ostriker, P. J. Steinhardt,
 Sci.Am. {\bf 284N1} 46, (2001). 

\bibitem{dgp}
G. Dvali, G. Gabadadze, M. Porrati,  Phys. Lett. {\bf B485}, 208 (2000) 
[hep-th/0005016]; 
G. Dvali, G. Gabadadze, M. Shifman, Phys. Lett. {\bf B497}, 271 (2001) 
[hep-th/0010071]. 

\end{thebibliography}
\end{document}